\documentclass[journal]{IEEEtran}
\usepackage{cite}
\usepackage{multirow}
\ifCLASSINFOpdf
  \usepackage[pdftex]{graphicx}
  \graphicspath{{figure/}}
  \DeclareGraphicsExtensions{.pdf,.jpeg,.png}
\else
  \usepackage[dvips]{graphicx}
  \graphicspath{{figure/}}
  \DeclareGraphicsExtensions{.eps}
\fi

\usepackage{makecell}
\usepackage{amsmath}
\usepackage{url}
\usepackage{amssymb}
\usepackage{bm}
\usepackage{epstopdf}
\usepackage{subfigure}
\begin{document}
\title{A Novel Speech Feature Fusion Algorithm for Text-Independent Speaker Recognition}

\author{Biao~Ma,~Chengben~Xu,~and~Ye~Zhang
\thanks{This work was supported by the National Natural Science Foundation of China under Grant 61866024. \emph{(Corresponding author: Ye Zhang.)}}
\thanks{The authors are with the Department of Electronic and Information Engineering, Nanchang University, Nanchang 330031, China (e-mail: zhangye@ncu.edu.cn).}
}


\maketitle

\begin{abstract}
A novel speech feature fusion algorithm with independent vector analysis (IVA) and parallel convolutional neural network (PCNN) is proposed for text-independent speaker recognition.
Firstly, some different feature types, such as the time domain (TD) features and the frequency domain (FD) features, can be extracted from a speaker's speech,
and the TD and the FD features can be considered as the linear mixtures of independent feature components (IFCs) with an unknown mixing system.
To estimate the IFCs, the TD and the FD features of the speaker's speech are concatenated to build the TD and the FD feature matrix, respectively.
Then, a feature tensor of the speaker's speech is obtained by paralleling the TD and the FD feature matrix.
To enhance the dependence on different feature types and remove the redundancies of the same feature type,
the independent vector analysis (IVA) can be used to estimate the IFC matrices of TD and FD features with the feature tensor.
The IFC matrices are utilized as the input of the PCNN to extract the deep features of the TD and FD features, respectively.
The deep features can be integrated to obtain the fusion feature of the speaker's speech.
Finally, the fusion feature of the speaker's speech is employed as the input of a deep convolutional neural network (DCNN) classifier for speaker recognition.
The experimental results show the effectiveness and performances of the proposed speaker recognition system.
\end{abstract}


\IEEEpeerreviewmaketitle

\section{Introduction} \label{sec:introduction}
\IEEEPARstart{S}{peaker} recognition is to identify the speaker identity by speakers' voices, \cite{1}, \cite{3}, \cite{4}.
Generally, the speaker recognition can be categorized into text-dependent and text-independent recognition tasks.
A text-dependent speaker recognition system attempts to recognize a speaker by a pre-defined phrase,
and a text-independent speaker recognition system does not expect the speaker to speak a particular phrase \cite{liu2015deep} \cite{5}.
In this paper, we focus on the text-independent speaker recognition system and some methods have been proposed for text-independent speaker recognition.
In \cite{7}, the Gaussian mixture model was built with the mel-frequency cepstral coefficients (MFCCs) for text-independent speaker identification.
In \cite{8352546}, a framework based on the triplet loss and a deep convolutional neural network architecture was trained with the spectrogram features or fbank features for text-independent speaker verification.
In \cite{XU2020394}, a network with residual connections and squeeze-and-excitation attention was trained with three losses and spectrograms for text-independent speaker verification.
Using the log-mel filterbank features,
two DCNNs, i.e., residual neural network (ResNet) and visual geometry group (VGG) nets, with a self-attention (SA) mechanism
were proposed for text-independent speaker identification \cite{15}.
In \cite{BIAN201959}, a ResNet with SA mechanism was trained with the Cluster-Range Loss and the log Fbank coefficients for text-independent speaker recognition.
However, only one feature type was utilized for the speaker recognition in all above-mentioned methods.
Generally, the performances of a speaker recognition system can be improved with the fusion feature by combining several different feature types.
Recently, some feature fusion methods have been proposed for speaker recognition.
In \cite{li2019speaker}, the MFCCs and log-mel filterbank feature were incorporated by a multi-feature integration method for speaker verification.
In \cite{li20f_interspeech}, the MFCCs and perceptual linear predictives, extracted from the same frame of the speaker's speech,
were integrated to obtain the integrated features for speaker verification.
In \cite{abraham2021deep}, the fusion feature of the speaker' speech was obtained by concatenating the MFCCs and its first-order derivatives and the chroma energy normalized statistics (CENS) features,
a convolutional neural network (CNN) was trained with the fusion features for speaker identification.
In \cite{17}, the LPCs and MFCCs with their first order delta coefficients were combined with a dilated 1D convolutional filter to obtain a frame-level embedding,
and then the utterance-level embedding was obtained by aggregating the frame-level embeddings with the average pooling for speaker verification.
In \cite{20}, the fusion feature of the speaker'speech was obtained with the concatenation of the i-vectors of the speaker's speech for speaker identification.

In this paper, we propose a novel speech feature fusion algorithm to obtain the fusion feature of the TD and FD features of the speaker's speech signal for text-independent speaker recognition.
First, the TD and FD features, such as linear predictive codings (LPCs) and MFCCs, are extracted from a speaker's speech,
and the TD and the FD features can be considered as the linear mixtures of IFCs with an unknown mixing system.
To estimate the IFCs, the TD and the FD features of the speaker's speech are concatenated to build the TD and the FD feature matrix, respectively.
Then, a feature tensor is obtained by paralleling these TD and FD feature matrix.
To enhance the dependence on different feature types and remove the redundancies of the same feature type,
the IVA can be utilized to estimate the IFC matrices of the TD and FD features with the feature tensor,
and the demixing tensor can be regarded as the speaker model.
The IFC matrices of the TD and FD features are used as the input of the PCNN to extract the deep features of the TD and FD features, respectively.
The fusion feature of the speaker's speech can be obtained by integrating the deep features.
Finally, the fusion feature is utilized as the input of the DCNN classifier for speaker recognition.

The contributions of this study can be summarized as three aspects.
First, a feature mixing model is introduced in this paper, i.e., the TD and the FD features can be considered as the linear mixtures of IFCs with an unknown mixing system.
Second, we propose a novel speech feature fusion algorithm to fuse the speech's TD and FD features for text-independent speaker recognition.
Generally, for the TD and FD features, such as LPCs and MFCCs, there are some complementarities between the TD and FD features of the speaker's speeches.
The LPCs are based on a theory of the speech production mechanism while the MFCCs are based on the speech perception by the human auditory system.
However, there may be some redundancies for the same type of the speech features, which may decrease the performances of the speaker recognition system.
The IVA can be utilized to extract the IFCs of the same speech feature type to remove their redundancies,
and also enhance the dependence on the different speech feature types for improving the performances of the speaker recognition system.
Third, a novel approach for building the speaker model for speaker recognition is proposed by estimating the demixing tensor.
In the IVA, the demixing tensor is formed individually for each speaker,
and it can be employed as the speaker model to obtain the estimation of the IFC matrices of the TD and FD features, respectively.

The outline structure of this paper can be organized as follows.
The related works are addressed in Section \ref{sec:relatedwork}.
The proposed speaker recognition system is detailed in the Section \ref{sec:propposedsry}.
The Section \ref{sec:exp} presents the experiments.
The Section \ref{sec:conclusion} concludes this study.

\section{Related Works} \label{sec:relatedwork}
\subsection{TD and FD Features}
Generally, the speech signal changes continuously, and it is unstationary.
The speech signal can be divided into many frames, and the duration of which is 20 to 30 millisecond (ms).
During this interval, the speech signal is assumed to be stationary and the TD and FD features can be extracted from these frames.
The TD features are calculated from the frames of the raw speeches in time-domain,
such as short-term energy, short-term magnitude, short-time zero-crossing rate, short-term auto correlation \cite{6557272},
LPCs \cite{1451722}, linear predictive cepstral coefficients (LPCCs) \cite{rao2015language}, etc.
The FD features are calculated in frequency-domain by using the Fourier transform to convert the speech signals from time-domain to frequency-domain,
such as MFCCs, log-magnitude spectral feature (LOG-MAG),
log-mel filterbank feature (LOG-MEL) \cite{7887742}, perceptual linear prediction \cite{Hermansky1990Perceptual},
gammatone frequency cepstral coefficients (GFCCs) \cite{4517928}, power-normalized cepstral coefficients (PNCCs) \cite{7439789}, etc.

In this paper, the TD and FD features focus on the LPCs and MFCCs for speaker recognition, respectively.
The voice activity detection is used to remove silence and unvoiced sounds in the original speech signals.
Then, the speech signals are pre-emphasized with a pre-emphasis coefficient of 0.97.
The pre-emphasized signals are divided into overlapping frames with a frame-length of $U$ and a frame-shift of $V$, and each frame is multiplied by a Hamming window.

\subsubsection{LPCs Extraction}
In the all-pole filter model, a speech sample $q_t(u)$ is assumed to be a linear combination of $R$ past samples and an error $e_t(u)$ for the $t^{\text{th}}$ frame.
\begin{equation}
  \begin{array}{*{20}{c}}
    {{q_t}(u) = \sum\limits_{r = 1}^R {{o_{rt}}{q_t}(u - r) + {e_t}(u)} }\\
    {t = 1, \ldots ,T;u = 1, \ldots ,U}
    \end{array}
  \label{lpc}
\end{equation}
where $q_t(u)$ is the $u^{\text{th}}$ sample of the $t^{\text{th}}$ frame of the speech signal, $o_{rt}$ is a filter coefficient (LPC) of order $r$, $e_t(u)$ is the error, $T$ is the number of frames.
The filter coefficients, i.e., LPCs, can be calculated by the least squares method.
The first order derivatives ($\Delta_{\text{l}}$) of the LPCs are calculated from the LPCs.
The second order derivatives ($\Delta^2_{\text{l}}$) of the LPCs are calculated from the first order derivatives of the LPCs.
For the $t^{\text{th}}$ frame, the LPCs, $\Delta_{\text{l}}$ and $\Delta^2_{\text{l}}$ are concatenated to form the TD feature vector ${\text{LPCs}}+\Delta_{\text{l}}+{\Delta^2_{\text{l}}}$.
For a speech signal, the LPCs matrix can be obtained with the ${\text{LPCs}}+\Delta_{\text{l}}+{\Delta^2_{\text{l}}}$.

\subsubsection{MFCCs Extraction}
The spectrograms of the frames are processed by the mel filterbank \cite{1163420}.
The LOG-MEL features are obtained with a log operation and the mel filterbank.
The MFCCs are calculated by applying the discrete cosine transform (DCT) \cite{1672377} to the LOG-MEL features.
The first order derivatives ($\Delta_{\text{m}}$) of the MFCCs can be calculated from the MFCCs.
The second order derivatives ($\Delta^2_{\text{m}}$) of the MFCCs can be calculated from the first order derivatives of the MFCCs.
For the $t^{\text{th}}$ frame, the MFCCs, $\Delta_{\text{m}}$ and $\Delta^2_{\text{m}}$ are concatenated to build FD feature vector the ${\text{MFCCs}}+\Delta_{\text{m}}+{\Delta^2_{\text{m}}}$.
For a speech signal, the MFCCs matrix is obtained with the ${\text{MFCCs}}+\Delta_{\text{m}}+{\Delta^2_{\text{m}}}$.

\subsection{Data Fusion With the IVA}
The IVA can be utilized for data fusion \cite{4176796}, \cite{28}.
In \cite{27}, an overview of some data fusion methods based on the independent component analysis (ICA) and IVA was presented
and the tradeoffs involved in the design of these fusion methods were also demonstrated.
Then, a new approach for fusion of disjoint subspaces was introduced for multimodal medical imaging data.
These multimodal medical imaging data, i.e., functional magnetic resonance imaging (MRI) and electroencephalography (EEG) data, were gathered from a group of the healthy controls and patients with the schizophrenia who performed an auditory oddball task.
In \cite{29}, the application of the joint ICA and transposed IVA model were considered for the fusion of the multimodal medical imaging data,
including functional MRI, structural MRI, and EEG data.
These medical imaging data were gathered from a group of healthy controls and patients with the schizophrenia who performed an auditory oddball task.

\begin{figure*}[!t]
  \centering
    \includegraphics[width=0.9\textwidth]{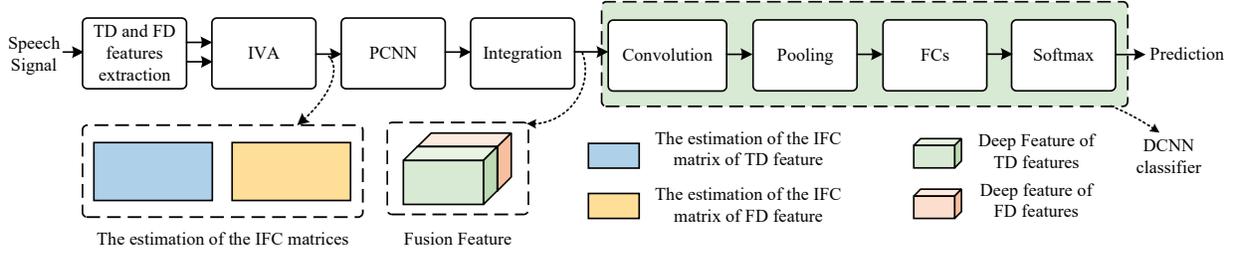}
  \caption{The block diagram of the proposed speaker recognition system. FCs denote the fully connected layers.}
  \label{sys}
\end{figure*}
\section{The Proposed Speaker Recognition System} \label{sec:propposedsry}
The block diagram of the proposed speaker recognition system is shown in Fig. \ref{sys}.
The TD and FD features, such as the LPCs and MFCCs, can be extracted from all frames of the speaker's speech to form a feature tensor.
The IVA can be used to estimate the IFC matrices of TD and FD features with the feature tensor,
and the demixing tensor obtained by the IVA can be regarded as the speaker's model.
The IFC matrices are utilized as the input of the PCNN to extract the deep features of the TD and FD features, respectively.
Then the fusion feature of the speaker's speech can be obtained by integrating the deep features,
and used as the input of the DCNN for speaker recognition.

\subsection{Estimation of the IFC Matrices by IVA} \label{feature and model}
\begin{figure}[!t]
  \centering
    \includegraphics[width=0.48\textwidth]{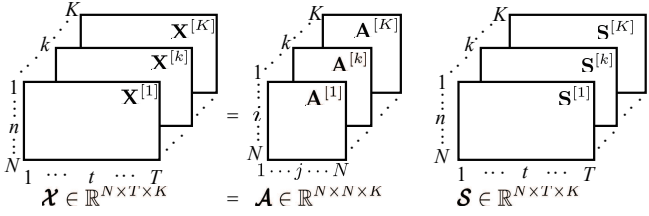}
  \caption{The feature mixing model.
  ${\pmb{\cal{X}}} \in \mathbb{R}^{N \times T \times K}$ denotes the feature tensor.
  ${\pmb{\cal{A}}} \in \mathbb{R}^{N \times N \times K}$ denotes the mixing tensor.
  ${\pmb{\cal{S}}} \in \mathbb{R}^{N \times T \times K}$ denotes the IFC tensor.
  }
  \label{mixing}
\end{figure}

Generally, some different feature types, such as the TD and the FD features, can be extracted from the same frame of a speaker's speech signal,
there is dependence for the different feature types, however,
there are redundancies for the same feature type.
The performances of a speaker recognition system may be improved by
enhancing the dependence on the different feature types and removing the redundancies of the same feature type.
In this paper, we introduce a feature mixing model, i.e.,
a feature vector can be considered as the linear mixtures of some unknown latent variables with an unknown mixing system, i.e.,
\begin{equation}
  {\bm{x}^{[k]}}(t) = {{\bf{A}}^{[k]}}{\bm{s}^{[k]}}(t)
  \label{Eq. 1}
\end{equation}
where ${\bm{x}^{[k]}}(t)={[x_1^{[k]}(t),\ldots,x_N^{[k]}(t)]^{\text{T}}} \in {\mathbb{R}^{N \times 1}}$ is the feature vector extracted from the $t^{\text{th}}$ frame for $k^{\text{th}}$ feature type, $t=1,\ldots,T$, $k=1,\ldots,K$,
$N$ denotes the feature dimension, $T$ denotes the number of the time frame and $K$ denotes the number of the feature type.
The superscript $\text{T}$ denotes transpose.
For the $k_1^{\text{th}}$ and $k_2^{\text{th}}$ feature type, there is dependence for ${\bm{x}^{[k_1]}}(t)$ and ${\bm{x}^{[k_2]}}(t)$, $k_1 \ne k_2$, $k_1,k_2=1,\ldots,K$.
For the $k^{\text{th}}$ feature type, there are redundancies for $x_{n'}^{[k]}(t)$ and $x_{n''}^{[k]}(t)$, $n' \ne n''$, $n',n'' = 1,\ldots,N$.
${{\bf{A}}^{[k]}}=[a_{ij}^{[k]}] \in{\mathbb{R}^{N\times N}}$ is an invertible mixing matrix, $i,j=1,\ldots,N$.
${\bm{s}^{[k]}}(t) = {[s_1^{[k]}(t),\ldots,s_N^{[k]}(t)]^{\text{T}}} \in {\mathbb{R}^{N \times 1}}$ are the unknown latent variables and cannot be directly observed.
In this paper, we assume that
the latent variables $s_n^{[k]}(t)$, $n=1,\ldots,N$,
are statistically mutually independent and called as the IFCs.
Then ${\bm{s}^{[k]}}(t)$ can be considered as the IFC vector.
For the $k_1^{\text{th}}$ and $k_2^{\text{th}}$ feature type, the ${\bm{s}^{[k_1]}}(t)$ and ${\bm{s}^{[k_2]}}(t)$ are dependent with each other, $k_1 \ne k_2$.
For the $k^{\text{th}}$ feature type, $s_{n'}^{[k]}(t)$ and $s_{n''}^{[k]}(t)$ are independent with each other, $n' \ne n''$.
For $k^{\text{th}}$ feature type, the feature matrix ${{\bf{X}}^{[k]}} = [{\bm{x}^{[k]}}(1),\ldots,{\bm{x}^{[k]}}(T)] \in \mathbb{R}^{N \times T}$ can be rewritten as
\begin{equation}
  {{\bf{X}}^{[k]}} = {{\bf{A}}^{[k]}}{{\bf{S}}^{[k]}}
  \label{Eq. 2}
\end{equation}
where ${{\bf{S}}^{[k]}} = [{\bm{s}^{[k]}}(1),\ldots,{\bm{s}^{[k]}}(T)] \in \mathbb{R}^{N \times T}$ is an IFC matrix.
The tensor $ \pmb{\cal{X}} \in \mathbb{R}^{N \times T \times K}$ is formed by paralleling $K$ feature matrix ${{\bf{X}}^{[k]}}$, then,
\begin{equation}
  {\pmb{\cal{X}}} = {\pmb{\cal{AS}}}
  \label{Eq. 4}
\end{equation}
where ${\pmb{\cal{A}}} \in \mathbb{R}^{N \times N \times K}$ is a mixing tensor, which is formed by paralleling $K$ mixing matrix ${{\bf{A}}^{[k]}}$.
The IFC tensor ${\pmb{\cal{S}}} \in \mathbb{R}^{N \times T \times K}$ is formed by paralleling $K$ IFC matrix ${{\bf{S}}^{[k]}}$, as shown in Fig. \ref{mixing}.

\begin{figure}[!t]
  \centering
    \includegraphics[width=0.48\textwidth]{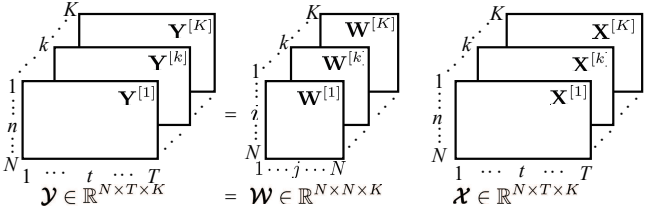}
  \caption{The IFC estimation model.
  ${\pmb{\cal{Y}}} \in \mathbb{R}^{N \times T \times K}$ denotes the estimation of the IFC tensor.
  ${\pmb{\cal{W}}} \in \mathbb{R}^{N \times N \times K}$ denotes the demixing tensor, which can be considered as speaker model.
  ${\pmb{\cal{X}}} \in \mathbb{R}^{N \times T \times K}$ denotes the feature tensor.
  }
  \label{iva}
\end{figure}

Generally, the tensor ${\pmb{\cal{X}}}$ can be used as a fusion feature and called as the feature tensor in this paper.
The feature tensor ${\pmb{\cal{X}}}$ can be considered as the linear mixtures of the IFC tensor ${\pmb{\cal{S}}}$ by with the mixing tensor ${\pmb{\cal{A}}}$.
The IFC tensor ${\pmb{\cal{S}}}$ can be considered as the fusion feature to improve the performances of the speaker recognition system.
In this paper, we propose using the IVA to estimate the IFC tensor ${\pmb{\cal{S}}}$, i.e.,
\begin{equation}
  {\pmb{\cal{Y}}} = {\pmb{\cal{WX}}}
  \label{Eq. 3}
\end{equation}
where ${\pmb{\cal{Y}}} \in  \mathbb{R}^{N \times T \times K}$ is the estimation of $\pmb{\cal{S}}$, and is formed by paralleling ${\bf{Y}}^{[k]}=[y_n^{[k]}(t)] \in \mathbb{R}^{N \times T}$, which is the estimation of ${\bf{S}}^{[k]}$, $k = 1,\ldots,K$, $n = 1,\ldots,N$, $t = 1,\ldots,T$.
${\pmb{\cal{W}}} \in  \mathbb{R}^{N \times N \times K}$ is the demixing tensor, which is formed by paralleling $K$ the demixing matrices ${{\bf{W}}^{[k]}}=[w_{ij}^{[k]}]\in \mathbb{R}^{N \times N}$,
and ${{\bf{W}}^{[k]}}$ is the estimation of the inverse of ${{\bf{A}}^{[k]}}$.
${\pmb{\cal{W}}}$ can be regarded as the speaker model since the demixing tensor is individually formed for each speaker.
The Eq. \ref{Eq. 3} can be regarded as the IFC estimation model, which is shown in Fig. \ref{iva}.

For the IVA, the mutual information minimization can be employed as the cost function to estimate ${\pmb{\cal{W}}}$ \cite{23}, i.e.,
\begin{equation}
  \begin{split}
    {I_{\text{IVA}}}
      &\buildrel \Delta \over = I[{{\bm{y}}_1(t)};\ldots;{{\bm{y}}_N(t)}]\\
      &= \sum\limits_{n = 1}^N {H[{{\bm{y}}_n(t)}]}  - H[{{\bm{y}}_1(t)},\ldots,{{\bm{y}}_N(t)}]\\
      &= \sum\limits_{n = 1}^N {H[{{\bm{y}}_n(t)}]}  - H[{{\bf{W}}^{[1]}}{{\bm{x}}^{[1]}(t)},\ldots,{{\bf{W}}^{[K]}}{{\bm{x}}^{[K]}(t)}]\\
      &= \sum\limits_{n = 1}^N {H[{{\bm{y}}_n(t)}]}  - \sum\limits_{k = 1}^K {\log |\det ({{\bf{W}}^{[k]}})|}  - {C_{\text{IVA}}}\\
      &= \sum\limits_{n = 1}^N {\left( {\sum\limits_{k = 1}^K {H[y_n^{[k]}(t)] - I[{{\bm{y}}_n}(t)]} } \right)} \\& \quad - \sum\limits_{k = 1}^K {\log |\det ({{\bf{W}}^{[k]}})|}  - {C_{\text{IVA}}}
  \end{split}
  \label{Eq. 5}
\end{equation}
where $I\left[\cdot\right]$ denotes the mutual information, $H\left[\cdot\right]$ denotes the entropy, and $\det \left( \cdot \right)$ denotes the determinant.
${\bm{y}}_n(t) = [y_n^{[1]}(t),\ldots,y_n^{[K]}(t)]^{\text{T}} \in {\mathbb{R}^{K \times 1}}$ is the estimation of $n^{\text{th}}$ source component vector (SCV) ${\bm{s}}_n(t) = [s_n^{[1]}(t),\ldots,s_n^{[K]}(t)]^{\text{T}} \in {\mathbb{R}^{K \times 1}}$ for the $t^{\text{th}}$ frame, $n\in\{1,\ldots,N\}$.
${C_{\text{IVA}}} = H[{{\bm{x}}^{[1]}(t)},\ldots,{{\bm{x}}^{[K]}(t)}]$ is a constant term.
When the cost function is minimized, the mutual information, $I[{{\bm{y}_n}(t)}]$, should be maximum, which means the dependence on different feature types is maximum.

Newton's method \cite{23} can be used for minimizing the IVA cost function to obtain the speaker model ${\pmb{\cal{W}}}$.
The gradient of the IVA cost function for ${\bf{w}}_n^{[k]}$ can be firstly calculated by
\begin{equation}
  \frac{{\partial {I_{{\text{IVA}}}}}}{{\partial {\bf{w}}_n^{[k]}}} =  E\left\{ {{\phi ^{[k]}}({{\bm{y}}_n(t)}){{\bm{x}}^{[k]}(t)}} \right\} - \frac{{{\bf{h}}_n^{[k]}}}{{{{\left( {{\bf{h}}_n^{[k]}} \right)}^{\text{T}}}{\bf{w}}_n^{[k]}}}
  \label{Eq. 11}
\end{equation}
where $E\left\{\cdot\right\}$ is the expectation.
${\bf{w}}_n^{[k]}=[w_{n1}^{[k]},\ldots,w_{nN}^{[k]}]^{\text{T}} \in {\mathbb{R}^{N \times 1}}$ denotes the $n^{\text{th}}$ row of the $k^{\text{th}}$ demixing matrix ${{\bf{W}}^{[k]}}$.
${\phi ^{[k]}}({{\bm{y}}_n(t)}) \buildrel \Delta \over =  - \partial \log p({{\bm{y}}_n(t)})/\partial y_n^{[k]}(t)$.
$p({{\bm{y}}_n(t)})$ is the joint probability density function (JPDF) of ${{\bm{y}}_n(t)}$.
${\bf{h}}_n^{[k]} \in \mathbb{R}^{N \times 1}$ is a unit-length vector, and it satisfies ${\bf{\tilde W}}_n^{[k]}{\bf{h}}_n^{[k]} = \bf{0}$, where ${\bf{\tilde W}}_n^{[k]} \in {\mathbb{R}^{(N - 1) \times N}}$ is the result of ${{\bf{W}}^{[k]}}$ removing the $n^{\text{th}}$ row \cite{33}, \cite{34}.
Then, the gradient of the IVA cost function for the $n^{\text{th}}$ demixing vector ${\bf{w}}_n = [{({\bf{w}}_n^{[1]})^{\text{T}}},\ldots,{({\bf{w}}_n^{[K]})^{\text{T}}}]^{\text{T}} \in \mathbb{R}^{KN \times 1}$ is calculated by
\begin{equation}
  \frac{{\partial {I_{\text{IVA}}}}}{{\partial {{\bf{w}}_n}}} = {\left[ {{{\left(\frac{{\partial {I_{\text{IVA}}}}}{{\partial {\bf{w}}_n^{[1]}}}\right)}^{\text{T}}},\ldots,{{\left(\frac{{\partial {I_{\text{IVA}}}}}{{\partial {\bf{w}}_n^{[K]}}}\right)}^{\text{T}}}} \right]^{\text{T}}}
  \label{Eq. 10}
\end{equation}


The Hessian matrix, ${{\bf{H}}_{{\text{SCV}}}} \buildrel \Delta \over = {\partial ^2}{I_{{\text{IVA}}}}/\partial {{\bf{w}}_n}\partial {\bf{w}}_n^{\text{T}}$, can be divided into $K \times K$ blocks and the dimension of each block is $N \times N$, i.e., ${{\bf{H}}_{{\text{SCV}}}} \in \mathbb{R}^{KN \times KN}$.
The rows of the $k_1^{\text{th}}$ block and the columns of the $k_2^{\text{th}}$ block of the Hessian matrix are denoted with ${{\bf{H}}_{k_1,k_2}} = {\partial ^2}{I_{{\text{IVA}}}}/\partial {\bf{w}}_n^{[{k_1}]}\partial {({\bf{w}}_n^{[{k_2}]})^{\text{T}}}\in \mathbb{R}^{N \times N}$, where  $k_1, k_2 = 1,\ldots, K$.
The off-block diagonal entries are
\begin{equation}
  {{\bf{H}}_{{k_1},{k_2}}} = E\left\{ {\frac{{\partial {\phi ^{[{k_2}]}}({{\bm{y}}_n(t)})}}{{\partial {\bf{w}}_n^{[{k_1}]}}}{{\left({{\bm{x}}^{[{k_2}]}(t)}\right)}^{\text{T}}}} \right\}, {k_1} \ne {k_2}
  \label{Eq. 14}
\end{equation}
The block diagonal entries of the Hessian matrix are
\begin{equation}
  {{\bf{H}}_{k,k}} = E\left\{ {\frac{{\partial {\phi ^{[k]}}({{\bm{y}}_n(t)})}}{{\partial {\bf{w}}_n^{[k]}}}{{\left({{\bm{x}}^{[k]}(t)}\right)}^{\text{T}}}} \right\} + \frac{{{\bf{h}}_n^{[k]}{{\left({\bf{h}}_n^{[k]}\right)}^{\text{T}}}}}{{{{\left(\left({\bf{h}}_n^{[k]}\right)^{\text{T}}{\bf{w}}_n^{[k]}\right)}^2}}}
  \label{Eq. 13}
\end{equation}
The ${{\bf{w}}_n^{\text{new}}}$ can be obtained with
\begin{equation}
  {{\bf{w}}_n^{\text{new}}} \leftarrow {{\bf{w}}_n^{\text{old}}} - \eta {\bf{H}}_{\text{SCV}}^{ - 1}\frac{{\partial {I_{\text{IVA}}}}}{{\partial {{\bf{w}}_n}}}
  \label{Eq. 15}
\end{equation}
where $\eta$ is the learning rate.

In this paper, the JPDF of the $n^{\text{th}}$ estimated SCV is considered as a zero-mean and real-valued $K$-dimensional multivariate Gaussian distribution
\begin{equation}
  p({{\bm{y}}_n}(t)|{{\bf{\Psi }}_n}) = \frac{1}{{{{(2\pi )}^{\frac{K}{2}}}\det {{({{\bf{\Psi }}_n})}^{\frac{1}{2}}}}}\exp \left( { - \frac{1}{2}{{\bm{y}}_n}{{(t)}^{\text{T}}}{\bf{\Psi }}_n^{ - 1}{{\bm{y}}_n}(t)} \right)
  \label{Eq. 16}
\end{equation}
where the estimation of the SCV covariance matrix ${ {\bf{\Psi }}_n}$ can be calculated with maximum likelihood estimates.
\begin{equation}
  {\hat {\bf{\Psi }}_n} = \frac{1}{T}\sum\limits_{t = 1}^{T} {{{\bm{y}}_{{n}}}(t){\bm{y}}_{{n}}^{\text{T}}(t)}
  \label{Eq. 17}
\end{equation}
The ${\phi ^{[k]}}({{\bm{y}}_n(t)})$ can be calculated by
\begin{equation}
  {\phi ^{[k]}}({{\bm{y}}_n(t)}) = {\{\hat {\bf{\Psi }}_n^{ - 1}{{\bm{y}}_n(t)}\} _k}
  \label{Eq. 18}
\end{equation}
where $\{\cdot\}_k$ denotes the index.
Then, the Eq. \ref{Eq. 14} and Eq. \ref{Eq. 13} can be simplified because
\begin{equation}
  E\left\{ {\frac{{\partial {\phi ^{[{k_2}]}}\left({{\bm{y}}_n(t)}\right)}}{{\partial {\bf{w}}_n^{[{k_1}]}}}{{\left({{\bm{x}}^{[{k_2}]}(t)}\right)}^{\text{T}}}} \right\} = {\left\{ {\hat {\bf{\Psi }}_n^{ - 1}} \right\}_{{k_1},{k_2}}}{\bf{R}}_x^{[{k_1},{k_2}]}
  \label{Eq. 21}
\end{equation}
where $\{\cdot\}_{k_1,k_2}$ denotes index.
The estimation of ${\bf{R}}_x^{[{k_1},{k_2}]}$ is obtained by
\begin{equation}
  {\bf{\hat R}}_x^{[{k_1},{k_2}]} = \frac{1}{T}\sum\limits_{t = 1}^{T} {{{\bm{x}}^{[{k_1}]}}(t){{\left( {{{\bm{x}}^{[{k_2}]}}(t)} \right)}^{\text{T}}}}
  \label{Rx_pr}
\end{equation}
\begin{figure}[!t]
  \small
  \centering
    \includegraphics[width=0.4\textwidth]{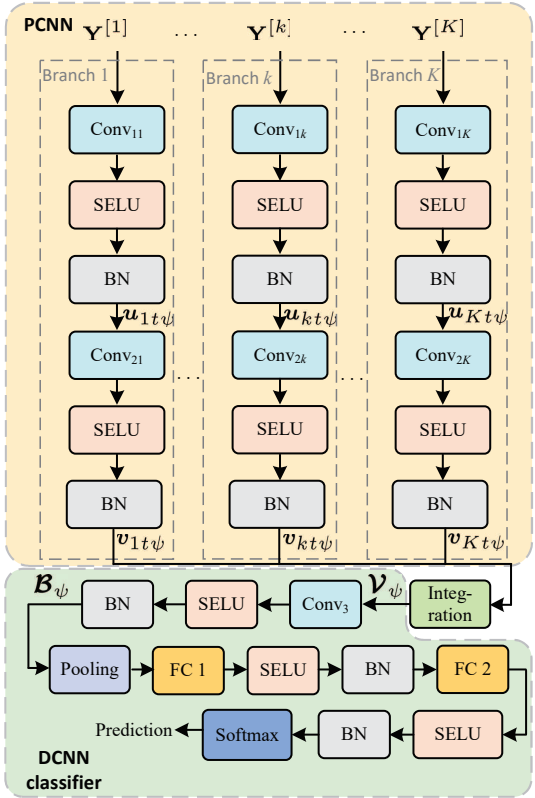}
  \caption{The block diagram of the PCNN-I. `I' denotes the capital of the first letter of the \textbf{i}ntegration.
  The PCNN-I is composed of PCNN and DCNN classifier.
  SELU denotes the scaled exponential linear units.
  FC denotes the fully connected layer. BN denotes the batch normalization.
  }
    \label{pcnn-i}
\end{figure}

\subsection{Fusion Feature}
For $k^{\text{th}}$ branch of the PCNN in Fig. \ref{pcnn-i}, the ${{{\bm{y}}^{[k]}}\left( t \right)}$ of the $t^{\text{th}}$ frame of $\psi^{\text{th}}$ sentence is firstly convoluted by $C_1$ convolutional kernels with the kernel size of $n_1 \times 1$.
The outputs of the first convolutional layer are activated with the the scaled exponential linear units (SELU) \cite{SELU}.
The activated outputs of the first convolutional layer are convoluted by $C_2$ convolutional kernels with kernel size of $n_2 \times 1 \times C_1$,
and they are also activated with the SELU, i.e.,
\begin{equation}
  {{\bm{u}}_{kt\psi}} = {\text{BN}}\left({\text{SELU}}\left( {{\text{Conv}_{1k}}\left( {{{\bm{y}}^{[k]}}\left( t; \psi \right);{{\bf{\Phi}} _{1k}}} \right)} \right)\right)
  \label{pcnn1 eq}
\end{equation}
\begin{equation}
  {{\bm{v}}_{kt\psi}} = {\text{BN}}\left({\text{SELU}}\left( {{\text{Conv}_{2k}}\left( {{{\bm{u}}_{kt\psi}};{{\bf{\Phi}} _{2k}}} \right)} \right)\right)
  \label{pcnn2 eq}
\end{equation}
where ${{\bm{u}}_{kt\psi}}\in {\mathbb{R}^{(N-n_1+1) \times 1 \times C_1}}$ denotes the outputs of the first convolutional layer of the $t^{\text{th}}$ frame of the $\psi^{\text{th}}$ sentence.
${{\bm{v}}_{kt\psi}}\in {\mathbb{R}^{(N-n_1-n_2+2) \times 1 \times C_2}}$ denotes the outputs of the second convolutional layer of the $t^{\text{th}}$ frame of the $\psi^{\text{th}}$ sentence.
${\text{BN}}\left(  \cdot  \right)$ denotes the batch normalization \cite{ioffe2015batch}.
${\text{SELU}}\left(  \cdot  \right)$ denotes the SELU non-linear activation function.
${\text{Conv}_{1k}}\left( { \cdot ;{{\bf{\Phi}} _{1k}}} \right)$ and ${\text{Conv}_{2k}}\left( { \cdot ;{{\bf{\Phi}} _{2k}}} \right)$ denote the first convolutional operation with the parameters ${\bf{\Phi}} _{1k}$ and the second convolutional operation with the parameters ${\bf{\Phi}} _{2k}$, respectively.
The deep feature of the $k^{\text{th}}$ feature type can be obtained by concatenating all outputs of the second convolutional layer, i.e.,
\begin{equation}
  \bm{V}_{k\psi} = \left[ {{\bm{v}_{k1\psi }}, \ldots ,{\bm{v}_{kT\psi }}} \right]
  \label{segment feat map pcnn}
\end{equation}
where $\bm{V}_{k\psi} \in \mathbb{R}^{(N-n_1-n_2+2) \times T \times C_2}$ denotes the deep feature.
The fusion feature of the speaker's $\psi^{\text{th}}$ sentence can be obtained by integrating $K$ deep features, i.e,
\begin{equation}
  {\pmb{{\cal V}}_\psi } = \left\{ {{\bm{V}_{1\psi }}, \ldots ,{\bm{V}_{K\psi }}} \right\}
  \label{parallel pcnn}
\end{equation}
where $\pmb{{\cal V}}_\psi \in {\mathbb{R}^{(N-n_1-n_2+2) \times T \times C_2K}}$ denotes the fusion feature of the speaker's $\psi^{\text{th}}$ sentence.
The fusion feature is convoluted with $C_3$ convolutional kernels, and the size of which is $n_3 \times 1 \times C_2K$ with the dilation of $D$.
The outputs of the third convolutional layer are activated with the the SELU.
\begin{equation}
  {\pmb{{\cal B}}_\psi } = {\text{BN}}\left( {{\text{SELU}}\left( {{\text{Conv}_3}\left( {{\pmb{{\cal V}}_\psi };{{\bf{\Phi }}_3}} \right)} \right)} \right)
  \label{conv1}
\end{equation}
where ${\pmb{{\cal B}}_\psi } \in {\mathbb{R}^{\left[N-n_1-n_2-n_3-(D-1)(n_3-1)+3\right] \times T \times C_3}}$ denotes the output of the third convolutional layer.
${\text{Conv}}{_3}( \cdot ;{{\bf{\Phi }}_3})$ denotes the third convolution with the parameters ${{\bf{\Phi }}_3}$.
A statistics pooling is applied on the ${\pmb{{\cal B}}_\psi }$ to compute its mean vector and variance vector over $t$, i.e.,
\begin{equation}
  {{\bm{\varphi }}_{\psi}} = \mathop {{\text{mean}}}\limits_t \left( {\pmb{{\cal B}}_\psi} \right)
  \label{pool_mean 2 eq}
\end{equation}
\begin{equation}
  {{\bm{\gamma }}_{\psi}} = \mathop {{\text{var}}}\limits_t \left( {\pmb{{\cal B}}_\psi} \right)
  \label{pool_var 2 eq}
\end{equation}
where ${{\bm{\varphi }}_{\psi}} \in {\mathbb{R}^{C_3\left[N-n_1-n_2-n_3-(D-1)(n_3-1)+3\right] \times 1}}$
and ${{\bm{\gamma }}_{\psi}} \in {\mathbb{R}^{C_3\left[N-n_1-n_2-n_3-(D-1)(n_3-1)+3\right] \times 1}}$ denote mean vector and variance vector, respectively.
A segment embedding is obtained by concatenating the average vector and variance vector, respectively, i.e.,
\begin{equation}
  {{\bm{e}}_\psi } = \left[ {{{\bf{\varphi }}_{\psi }};{{\bm{\gamma }}_{\psi }}} \right]
  \label{cat_embedding 2 eq}
\end{equation}
where ${\bm{e}_\psi} \in {\mathbb{R}^{2C_3\left[N-n_1-n_2-n_3-(D-1)(n_3-1)+3\right] \times 1}}$ denotes the segment embedding.
The segment embedding ${\bm{e}_\psi}$ is used as the input of the fully connected (FC) layers, i.e.,
\begin{equation}
  {\bm{m}_\psi} = {\text{BN}}\left({\text{SELU}}\left( {{{\bm{W}}_1^{\text{T}}}{\bm{e}_\psi} + {{\bm{c}}_1}} \right)\right)
  \label{fc1 2 eq}
\end{equation}
\begin{equation}
  {\bm{n}_\psi} = {\text{BN}}\left({\text{SELU}}\left( {{{\bm{W}}_2^{\text{T}}}{\bm{m}_\psi} + {{\bm{c}}_2}} \right)\right)
  \label{fc2 2 eq}
\end{equation}
where ${\bm{m}_\psi}\in \mathbb{R}^{F_1 \times 1}$ and ${\bm{n}_\psi} \in \mathbb{R}^{F_2 \times 1}$ are the outputs of the first and second FC layer.
${\bm{W}}_1 \in \mathbb{R}^{2C_3\left[N-n_1-n_2-n_3-(D-1)(n_3-1)+3\right] \times F_1}$ and ${\bm{W}}_2 \in \mathbb{R}^{F_1 \times F_2}$ are the weights of the first and second FC layers, respectively.
${\bm{c}}_1 \in \mathbb{R}^{F_1 \times 1}$ and ${\bm{c}}_2 \in \mathbb{R}^{F_2 \times 1}$ are the biases of the first and second FC layers, respectively.
Finally, the outputs of the second FC layer are used as the input of the softmax layer, i.e.,
\begin{equation}
  {o_{\lambda\psi} } = \frac{{\exp \left( {{\bm{w}}_\lambda ^{\text{T}}{\bm{n}_\psi} + {b_\lambda }} \right)}}{{\sum\limits_\tau  {\exp \left( {{\bm{w}}_\tau ^{\text{T}}{\bm{n}_\psi} + {b_\tau }} \right)} }}
  \label{output_sm eq}
\end{equation}
where the $o_{\lambda\psi}$ denotes the probability that the $\psi^{\text{th}}$ sentence is predicted to be the $\lambda^{\text{th}}$ speaker.
${\bm{w}}_\lambda \in \mathbb{R}^{F_2 \times 1}$ and ${\bm{w}}_\tau \in \mathbb{R}^{F_2 \times 1}$ denote the $\lambda^{\text{th}}$ and $\tau^{\text{th}}$ column weights in the softmax layer, respectively.
$b_\lambda$ and $b_\tau$ are scalars, and they denote the biases of the $\lambda^{\text{th}}$ and $\tau^{\text{th}}$ neuron, respectively.
$\lambda \in \left\{ {1, \ldots ,{C_{{\text{spk}}}}} \right\}$, $\tau=1,\ldots,{C_{{\text{spk}}}}$.
$C_{{\text{spk}}}$ denotes the number of the speakers.
The cross entropy loss is used to optimize the PCNN-I.
\begin{equation}
  {L_{{\text{CE}}}} = - \frac{1}{\Gamma }\sum\limits_{\psi  = 1}^\Gamma  {\sum\limits_{\lambda  = 1}^{{C_{{\text{spk}}}}} {{l_{\lambda \psi }}\log {o_{\lambda \psi }}} }
  \label{ce eq}
\end{equation}
where ${L_{{\text{CE}}}}$ denotes cross entropy loss.
$\Gamma$ denotes the number of the sentences of the mini-batch.
\begin{equation}
  {l_{\lambda \psi }} = \left\{ {\begin{array}{*{20}{c}}
    {1,\quad \lambda  = \psi }\\
    {0,\quad \lambda  \ne \psi }
    \end{array}} \right.
  \label{label eq}
\end{equation}
where $l_{\lambda \psi }$ denotes the speaker labels.

\section{Experiments} \label{sec:exp}
\subsection{Datasets}
Two datasets are employed to evalute the performances of the proposed speaker recognition system, i.e.,
the Free ST Chinese Mandarin Corpus\footnote{[Online]. Available: \url{http://www.openslr.org/38}} (FSCMC),
AISHELL-1 \cite{8384449}.
The FSCMC is an open-source Mandarin speech corpus, and it is recorded in silence in-door environment using cellphone.
The sampling rate of the speech signals in this dataset is 16 kHz, the bit depth of the speech signals is 16 bits, and the speech signals are mono.
This corpus contains 855 speakers with 443 males and 412 females, and each speaker has 120 utterances.
All utterances are carefully transcribed and checked by human.
The AISHELL-1 is also an open-source Mandarin speech corpus.
It includes 400 speakers over 170 hours of Mandarin speech data, the gender is balanced with 47\% male and 53\% female, and most speakers are of age 16 to 25.
The audio utterances are resampled to 16 kHz and 16-bit WAV format.

\subsection{Speech Features}
For each sentence, the silence and unvoiced sounds of the training and testing sentences are removed with the voice activation detection.
The duration of the training and testing sentences is fixed at 3.015 s.
If the duration of the training and testing sentences is longer than 3.015 s, the segments with the duration of 3.015 s are randomly selected from these sentences.
Otherwise, these sentences are padded into 3.015 s with themselves.
The training and testing sentences are pre-emphasized with a pre-emphasis of 0.97.
The pre-emphasized speech signals are divided into overlapping frames with a frame-length of 25 ms and a frame-shift of 10 ms,
so the number of the frame-length and the frame-shift are $U=400$ and $V=160$ (because sample rate of the speech signals is 16 kHz), respectively.
Each frame of the speaker's speech is multiplied by a Hamming window.
For each sentence, there are $T=300$ frames.

\subsubsection{LPCs Matrix}
For each frame of the speaker's speech, the LPCs are calculated with Eq. \ref{lpc} and the order $R$ is set to 13.
The first derivatives $\Delta_{\text{l}}$ and the second derivatives $\Delta^2_{\text{l}}$ of the LPCs are calculated to capture the information about how the LPCs changes over time.
The LPCs, $\Delta_{\text{l}}$ and $\Delta^2_{\text{l}}$ are concatenated to form TD feature vector LPCs+$\Delta_{\text{l}}$+${\Delta }^2_{\text{l}} \in \mathbb{R}^{39 \times 1}$, i.e., $N=39$.
For each sentence, the LPCs matrix ${{\bf{X}}^{[1]}} \in \mathbb{R}^{39 \times 300}$ can be obtained by concatenating all TD feature vector.

\subsubsection{MFCCs Matrix}
For each frame of the speaker's speech, the spectrograms are obtained with discrete fourier transform, which are then processed by a mel filterbank of 39 triangular filters.
The LOG-MEL features are obtained by the log operation for the results of the mel filterbank.
The 13-dimensional MFCCs are calculated by applying DCT to LOG-MEL features.
The first derivatives $\Delta_{\text{m}}$ and second derivatives $\Delta^2_{\text{m}}$ of the MFCCs are calculated to capture the dynamic changes of the MFCCs over time.
The MFCCs, $\Delta_{\text{m}}$ and $\Delta^2_{\text{m}}$ are concatenated to obtain FD feature vector MFCCs+$\Delta_{\text{m}}$+$\Delta^2_{\text{m}} \in \mathbb{R}^{39 \times 1}$.
For each sentence, the MFCCs matrix ${{\bf{X}}^{[2]}} \in \mathbb{R}^{39 \times 300}$ is obtained by concatenating all FD feature vector.

\subsubsection{Feature Tensor}
For a sentence of the speaker,
the feature tensor ${\pmb{\cal{X}}}$ can be obtained by paralleling ${{\bf{X}}^{[1]}}$ and ${{\bf{X}}^{[2]}}$, i.e., $K=2$.
It is worth noting that the feature tensor ${\pmb{\cal{X}}} \in \mathbb{R}^{39 \times 300 \times 2}$ is also the fusion feature without using the IVA.
\begin{table}[!t]
  \renewcommand{\arraystretch}{1.3}
  \caption{The PCNN-I Structure}
  \label{pcnn-i table}
  \centering
    \begin{tabular}{cccc}
    \hline
    No. & Layer Name & Kernel Size & Output\\
    \hline
    \multirow{2}{*}[0ex]{1} & ${{\bf{Y}}^{[1]}}$  & - & (39, 300, 1)\\
                        & ${{\bf{Y}}^{[2]}}$ & - &(39, 300, 1)\\
    \hline
    \multirow{2}{*}[0ex]{2} & ${\text{Convolution}}_{11}$ & $n_1 \times 1$ & (40$-n_1$, 300, 32)\\
                       & ${\text{Convolution}}_{12}$ & $n_1 \times 1$ & (40$-n_1$, 300, 32)\\
    \hline
    \multirow{2}{*}[0ex]{3} & ${\text{Convolution}}_{21}$ & $5 \times 1 \times 32$ & (36$-n_1$, 300, 32)\\
                       & ${\text{Convolution}}_{22}$ & $5 \times 1 \times 32$ & (36$-n_1$, 300, 32)\\
    \hline
    \multirow{1}{*}[0ex]{4} & Integration & - & (36$-n_1$, 300, 64)\\
    \multirow{1}{*}[0ex]{5} & ${\text{Convolution}}_{3}$ & $7 \times 1 \times$ 64 & (30$-n_1$, 300, 64)\\
    \multirow{1}{*}[0ex]{6} & Pooling & - & (3840$-$128$n_1$, 1)\\
    \multirow{1}{*}[0ex]{7} & FC 1 & - & (512, 1)\\
    \multirow{1}{*}[0ex]{8} & FC 2 & - & (512, 1)\\
    \multirow{1}{*}[0ex]{9} & Softmax & - & ($C_{\text{spk}}$, 1)\\
    \hline
    \multicolumn{4}{l}{\makecell[l]{$n_2=5$, $n_3=7$, $C_1=32$, $C_2=32$, $C_3=64$, $D=3$, \\$F_1=512$,$F_2=512$.}}
    \end{tabular}
\end{table}
\begin{table}[!t]
  \renewcommand{\arraystretch}{1.3}
  \caption{The NCNN Structure}
  \label{ncnn table}
  \centering
    \begin{tabular}{cccc}
    \hline
    No. & Layer Name & Kernel Size & Output\\
    \hline
    1 & ${\pmb{\cal{Y}}}$ & - & (39, 300, 2)\\
    2 & Convolution & $n \times K$ & (64, $T$)\\
    3 & Pooling & - & (128, 1)\\
    4 & FC 1 & - & (64, 1)\\
    5 & FC 2 & - & (64, 1)\\
    6 & Softmax & - & ($C_{\text{spk}}$, 1)\\
   \hline
    \end{tabular}
\end{table}
\begin{table}[!t]
  \renewcommand{\arraystretch}{1.3}
  \caption{The PCNN-C Structure}
  \label{pcnn-c table}
  \centering
    \begin{tabular}{cccc}
    \hline
    No. & Layer Name & Kernel Size & Output\\
    \hline
    \multirow{2}{*}[0ex]{1} &  ${{\bf{Y}}^{[1]}}$  & - & (39, 300, 1)\\
                        & ${{\bf{Y}}^{[2]}}$ & - &(39, 300, 1)\\
    \hline
    \multirow{2}{*}[0ex]{2} & ${\text{Convolution}}_{11}$ & $n \times 1$ & (40$-n$, 300, 32)\\
                       & ${\text{Convolution}}_{12}$ & $n \times 1$ & (40$-n$, 300, 32)\\
    \hline
    \multirow{2}{*}[0ex]{3} & ${\text{Convolution}}_{21}$ & $5 \times 1 \times 32$ & (36$-n$, 300, 32)\\
                       & ${\text{Convolution}}_{22}$ & $5 \times 1 \times 32$ & (36$-n$, 300, 32)\\
    \hline
    \multirow{2}{*}[0ex]{4} & Pooling 1 & - & (2304$-$64$n$, 1)\\
                       & Pooling 2 & - & (2304$-$64$n$, 1)\\
    \hline
    \multirow{1}{*}[0ex]{5} & Concatenation & - & (4608$-$128$n$, 1)\\
    \multirow{1}{*}[0ex]{6} & FC 1 & - & (512, 1)\\
    \multirow{1}{*}[0ex]{7} & FC 2 & - & (512, 1)\\
    \multirow{1}{*}[0ex]{8} & Softmax & - & ($C_{\text{spk}}$, 1)\\
    \hline
    \end{tabular}
\end{table}

\subsection{Performance Measure}
The accuracy (ACC) rate is employed to evaluate the performances of the speaker recognition system, i.e.,
\begin{equation}
  ACC = \frac{{number\;of\;correct\;classified\;samples}}{{number\;of\;total\;testing\;samples}} \times 100\%
  \label{ACC}
\end{equation}

\subsection{Baseline Systems}
In this section, we introduce some baseline speaker recognition systems \cite{15}, \cite{abraham2021deep}, \cite{17}, \cite{20}, \cite{chakroun2020efficient}.
In \cite{15}, the speech signals were pre-emphasized with a factor of 0.97.
The speech signals were divided into overlapping frames with a frame-length of 25 ms and a frame-shift of 10 ms, and each frame was multiplied with a Hamming window.
The log-mel filterbank features can be extracted from the frames of the speech signals, and then these features
were utilized as the input of two CNNs, i.e., ResNet and VGG, with multi-head SA mechanism for text-independent speaker identification.
For the sake of legibility, the ResNet and VGG with the multi-head mechanism were abbreviated as the ResNet + SA and VGG + SA, respectively.

In \cite{abraham2021deep}, the speech signals were pre-emphasized and segmented into frames with the length of 50 ms and a shift of 25 ms.
The 40-dimension MFCCs, its first-order derivatives and 12-dimension CENS features were computed from each frame,
and then these features were concatenated at frame level to yield a short-term feature vector, i.e., the fusion features.
A CNN was trained to identify the identity of the speaker with the MFCCs and its first-order derivatives, or the fusion features.
For the sake of legibility, the speaker recognition system with the MFCCs and its first-order derivatives was abbreviated as the MFCCs + CNN,
and the speaker recognition system with the fusion features was abbreviated as MFCCs + CENS + CNN.

In \cite{17}, the 20-dimensional LPCs and MFCCs and their first order delta coefficients were extracted from all frames of the speech signals for speaker verification.
The extracted features were normalized using cepstral mean and variance normalization (CMVN).
The LPCs and MFCCs with their first order delta coefficients, which were extracted from the same frame, were combined with a dilated 1D convolutional filter to obtain a frame-level embedding.
The utterance-level embedding was obtained by aggregating the frame-level embeddings across 200 frames using the average pooling.
For the sake of legibility, this speaker recognition system was abbreviated with LPCs-MFCCs + 1D-Triplet-CNN.

In \cite{20}, a speech signal of a speaker was segmented into overlapping frames and tapered with three different multitapers, i.e., thomson tapers (tt), sine tapers (st) and multipeak (mp) tapers.
The MFCCs and PNCCs were extracted from these frames to obtain six features, i.e., tt-MFCCs, st-MFCCs, mp-MFCCs, tt-PNCCs,  st-PNCCs, mp-PNCCs.
Then, the feature warping (FW) and CMVN were used to normalize these six features to obtain twelve features, i.e., tt-MFCCsFW, st-MFCCsFW, mp-MFCCsFW, tt-MFCCsCMVN, st-MFCCsCMVN, mp-MFCCsCMVN, tt-PNCCsFW, st-PNCCsFW, mp-PNCCsFW, tt-PNCCsCMVN, st-PNCCsCMVN, mp-PNCCsCMVN.
The i-vectors of the speaker's speech were calculated with these twelve features.
All i-vectors were fused by calculating the concatenation of these i-vectors for speaker identification by the extreme learning machine (ELM).
For the sake of legibility, this speaker recognition system was abbreviated with i-vector + ELM.

In \cite{chakroun2020efficient}, the speech signals were divided into several frames with the length of 25 ms and an overlap of 10 ms.
A 13-dimensional GFCC feature vector was extracted from a frame of a speaker's speech.
The mean and variance normalized GFCC (MVNGFCC) feature vector was obtained by calculating the mean and variance of a GFCC feature vector.
The i-vector of a sentence was calculated with all MVNGFCC feature vectors of a sentence,
and it was used as the input of the probabilistic linear discriminant analysis model for speaker identification.
For the sake of legibility, this speaker recognition system was abbreviated as the i-vector.

\subsection{Experiment 1}
One hundred speakers are randomly selected from the FSCMC and AISHELL-1.
For each speaker, 100 sentences are employed as the training data and another 20 sentences are used as the testing data.
The texts of the speech signals are different between the training data and the testing data.
For a sentence of a speaker, to obtain ${\bf{Y}}^{[1]} \in\mathbb{R}^{39 \times 300 \times 1}$ and ${\bf{Y}}^{[2]} \in\mathbb{R}^{39 \times 300 \times 1}$, the speaker model ${\pmb{\cal{W}}} \in\mathbb{R}^{39 \times 39 \times 2}$ is randomly initialized, and then ${\pmb{\cal{W}}}$ is estimated with ${\pmb{\cal{X}}}$ by using the IVA.
The learning rate $\eta$ of the IVA is initially set to 1.0, $\eta \ge 1 \times 10^{-6}$. $\eta$ is multiplied with 0.9 if the value of the Eq. \ref{Eq. 5} increased.
${\bf{Y}}^{[1]}$ and ${\bf{Y}}^{[2]}$ are calculated with ${\pmb{\mathcal{W}}}$ and ${\pmb{\cal{X}}}$,
and they are used as the input of the PCNN-I to extract the deep features of the LPCs matrix and MFCCs matrix, respectively.
The deep features can be integrated to obtain the fusion feature of the speaker's speech,
and then the fusion feature is used as the input of the DCNN classifier for speaker recognition.
The structure of the PCNN-I is shown in Table \ref{pcnn-i table}.
The PCNN-I is implemented with Pytorch \cite{paszke2017automatic},
and the Adam \cite{38} optimizer is used as the optimizer for the PCNN-I.
The proposed speaker recognition system is compared with the baseline speaker recognition systems \cite{15}, \cite{abraham2021deep}, \cite{17}, \cite{20}, \cite{chakroun2020efficient},
and the experimental schemes are consistent with the original paper.
The experimental results are shown in Table \ref{baseline},
where it can be seen that the proposed speaker recognition system achieves comparable performances compared with the baseline speaker recognition systems.
For example, on the FSCMC, the ACC of proposed speaker recognition system is 99.70\%,
and the ACC of the LPCs-MFCCs + 1D-Triplet-CNN \cite{17} is 99.65\%.
On the AISHELL-1, the ACC of the proposed speaker recognition system is 99.00\%,
and the ACC of the i-vector \cite{chakroun2020efficient} is 97.25\%.

\subsection{Experiment 2}
Two other speech feature fusion algorithms based on IVA are also introduced to compare the proposed speech feature fusion algorithm.
For the first fusion algorithm, ${{\bf{Y}}^{[1]}}$ and ${{\bf{Y}}^{[2]}}$ are paralleled to obtain the fusion feature of the speaker'speech,
and the fusion feature is used as the input of the non-parallel CNN (NCNN) for text-independent speaker recognition.
The SELU is utilized as the activation function of the convolution layer and two FC layers.
The structure of the NCNN is shown in Table \ref{ncnn table}.
For the second fusion algorithm, ${{\bf{Y}}^{[1]}}$ and ${{\bf{Y}}^{[2]}}$ are used as the input of the PCNN to obtain the deep features of the TD and FD features, respectively.
These deep features are pooled to obtain a mean vector and a variance vector, respectively.
The mean vector and the variance vector are concatenated to obtain the fusion feature of the speaker's speech.
The fusion feature is used as the input of the FC layers,
and output of the FC layers is utilized as the input of the softmax layer for text-independent speaker recognition.
The SELU is used as the activation function of the convolution layers in PCNN and two FC layers.
The activations of each layer is used as the input of the BN.
This feature fusion algorithm can be achieved with the PCNN-C,
where `C' denotes the capital of the first letter of the \textbf{c}oncatenation.
The structure of the PCNN-C is shown in Table \ref{pcnn-c table}.
The NCNN and PCNN-C are implemented with Pytorch \cite{paszke2017automatic},
and the Adam \cite{38} optimizer is used as the optimizer for the NCNN and PCNN-C, respectively.
The cross entropy loss is utilized as the loss function of the NCNN and PCNN-C, respectively.
The experimental results are shown in Table \ref{table pcnn} and Table \ref{table multi feature},
where it illustrates that the performances of the PCNN-I are better than the those of the NCNN and PCNN-C.
For example, across the Table \ref{table pcnn} and Table \ref{table multi feature},
for the kernel size of $3 \times 1$,
the ACC of ${{\bf{Y}}^{[1]}}$ + ${{\bf{Y}}^{[2]}}$ + PCNN-I is higher than the ACC of the ${{\bf{Y}}^{[1]}}$ + ${{\bf{Y}}^{[2]}}$ + PCNN-C by 0.70\%
and the ACC of ${{\bf{Y}}^{[1]}}$ + ${{\bf{Y}}^{[2]}}$ + PCNN-I is also higher than the ACC of ${\pmb{\cal{Y}}}$ + NCNN by 2.25\%.
The experimental results also demonstrate that the performances of the PCNN-C are better than the those of the NCNN.
For example, across the Table \ref{table pcnn} and Table \ref{table multi feature},
for the kernel size of $3 \times 1$,
the ACC of ${{\bf{Y}}^{[1]}}$ + ${{\bf{Y}}^{[2]}}$ + PCNN-C succeeds the ACC of the $\pmb{\cal{Y}}$ + NCNN by 1.85\%.

\subsection{Experiment 3}
To evaluate the performances of the speaker recognition system with different speech features,
100 speakers are selected from the FSCMC,
for each speaker, 100 sentences are used as the training data, and another 20 sentences are used as the testing data.
The texts of the speech signals are different between the training data and the testing data.
Four speech features, i.e., ${\pmb{\cal{Y}}}$, ${\pmb{\cal{X}}}$, ${\bf{X}}^{[1]}$, ${\bf{X}}^{[2]}$, are extracted from the sentence of the FSCMC,
and they are used as the input of the NCNN, respectively.
If ${\pmb{\cal{Y}}}$ or ${\pmb{\cal{X}}}$ is used as the input of the NCNN, $K$ is set to 2.
If ${\bf{X}}^{[1]}$ or ${\bf{X}}^{[2]}$ is used as the input of the NCNN, $K$ is set to 1.
The experimental results of ${\pmb{\cal{X}}}$ + NCNN and ${\pmb{\cal{Y}}}$ + NCNN are shown in Table \ref{table multi feature},
and the experimental results of ${{\bf{X}}^{[1]}}$ + NCNN and ${{\bf{X}}^{[2]}}$ + NCNN are shown in Table \ref{table single feature}.
From the Table \ref{table multi feature} and Table \ref{table single feature},
we can see that the speaker recognition system using multiple feature, i.e., ${\pmb{\cal{Y}}}$ or ${\pmb{\cal{X}}}$,
is superior to that of using single speech feature, i.e., ${\bf{X}}^{[1]}$ or ${\bf{X}}^{[2]}$.
For instance, compared with the ${{\bf{X}}^{[1]}}$ + NCNN with the kernel size of $3 \times 1$,
the ACC of the ${\pmb{\cal{X}}}$ + NCNN with the kernel size of $3 \times 2$ improves 6.35\%,
and the ACC of the ${\pmb{\cal{Y}}}$+NCNN with $3 \times 2$ increases 8.75\%.
The ACC of the speaker recognition system using ${\pmb{\cal{X}}}$ is also higher than that using single feature, i.e., ${{\bf{X}}^{[1]}}$ or ${{\bf{X}}^{[2]}}$.
For other kernel sizes, the NCNN using multiple feature also outperforms the NCNN using the single speech feature.
It can be attributed to the complementarities between the TD feature and FD feature of the speaker's speech.
From the Table \ref{table multi feature},
the ACC of the ${\pmb{\cal{Y}}}$ + NCNN is superior than that of the ${\pmb{\cal{X}}}$ + NCNN.
For instance, for kernel size $3 \times 2$,
the ACC of the ${\pmb{\cal{Y}}}$ + NCNN is higher than that of the ${\pmb{\cal{X}}}$ + NCNN by 2.40\%.
The reason for the improvements of the ACCs is that (a) the LPCs and MFCCs extracted from the same frame are complementary,
i.e., the LPCs are based on a theory of the speech production mechanism while the MFCCs are based on the speech perception by the human auditory system,
and (b) the dependence on the different feature types is enhanced and the redundancies of the same feature type are removed by using the IVA.
This experimental results support the benefit of the IVA, since the best performances of the speaker recognition system are obtained using ${\pmb{\cal{Y}}}$, thus approving our contributions.

\begin{table}[!t]
  \renewcommand{\arraystretch}{1.6}
  \caption{The ACCs (\%) of Different Speaker Recognition Systems. FSCMC denotes the Free ST Chinese Mandarin Corpus.}
  \label{baseline}
  \centering
    \begin{tabular}{cccc}
    \hline
    No.&Methods&FSCMC&AISHELL-1\\
    \hline
    1&$\bf{Y}^{\text{[1]}}$ + $\bf{Y}^{\text{[2]}}$ + PCNN-I ($3 \times 1$) & 99.70 & 99.00\\
    2&ResNet + SA \cite{15} & 94.95&89.15\\
    3&VGG + SA \cite{15} & 94.85 &88.20\\
    4&MFCCs + CNN \cite{abraham2021deep} & 93.60 &91.00\\
    5&MFCCs + CENS + CNN \cite{abraham2021deep} & 94.05 &92.20\\
    6&LPCs-MFCCs + 1D-Triplet-CNN \cite{17} &99.65 &99.75\\
    7&i-vector + ELM \cite{20} & 99.55&99.80\\
    8&i-vector \cite{chakroun2020efficient} &94.05&97.25\\
    \hline
    \end{tabular}
\end{table}
\begin{table}[!t]
  \renewcommand{\arraystretch}{1.6}
  \caption{The ACCs (\%) of the Speaker Recognition System Using $\bf{Y}^{\text{[1]}}$ and $\bf{Y}^{\text{[2]}}$ with the PCNN-C or PCNN-I on the FSCMC. The Kernel Size of the Second Convolution Layer is $5 \times 1$.}
  \label{table pcnn}
  \centering
    \begin{tabular}{ccccc}
    \hline
    Kernel Size ($n_1 \times 1$) & 1 $\times$ 1 & 3 $\times$ 1 & 5 $\times$ 1 & 7 $\times$ 1\\
    \hline
    $\bf{Y}^{\text{[1]}}$ + $\bf{Y}^{\text{[2]}}$ + PCNN-I & 99.45 & 99.70 & 99.55 & 99.45\\
    $\bf{Y}^{\text{[1]}}$ + $\bf{Y}^{\text{[2]}}$ + PCNN-C & 99.10 & 99.00 & 99.15 & 99.10\\
    \hline
    \end{tabular}
\end{table}
\begin{table}[!t]
  \renewcommand{\arraystretch}{1.6}
  \caption{The ACCs (\%) of the Speaker Recognition System Using ${\pmb{\cal{X}}}$ or ${\pmb{\cal{Y}}}$ with the different Kernel Sizes of the NCNN on the FSCMC.
  If ${\pmb{\cal{X}}}$ or ${\pmb{\cal{Y}}}$ is used as the input of the NCNN, $K$ is set to 2.
  NCNN denotes non-parallel CNN.}
  \label{table multi feature}
  \centering
    \begin{tabular}{ccccc}
    \hline
    Kernel Size ($n \times K$) & 1 $\times$ 2 &3 $\times$ 2 & 5 $\times$ 2 &7 $\times$ 2\\
    \hline
    ${\pmb{\cal{Y}}}$ + NCNN & 96.80 & 97.15 & 96.95 & 96.80\\
    ${\pmb{\cal{X}}}$ + NCNN & 94.45 & 94.75 & 95.15 & 95.60\\
    \hline
    \end{tabular}
\end{table}
\begin{table}[!t]
  \renewcommand{\arraystretch}{1.6}
  \caption{The ACCs (\%) of the Speaker Recognition System Using ${{\bf{X}}^{[1]}}$ or ${{\bf{X}}^{[2]}}$ with the Different Kernel Sizes of the NCNN on the FSCMC.
  If ${{\bf{X}}^{[1]}}$ or ${{\bf{X}}^{[2]}}$ is used as the input of the NCNN, $K$ is set to 1.
  NCNN denotes non-parallel CNN.}
  \label{table single feature}
  \centering
    \begin{tabular}{ccccc}
    \hline
    Kernel Size ($n \times K$) & 1 $\times$ 1 & 3 $\times$ 1 & 5 $\times$ 1 & 7 $\times$ 1\\
    \hline
    ${{\bf{X}}^{[1]}}$ + NCNN & 84.40  & 88.40 & 88.60 & 89.35\\
    ${{\bf{X}}^{[2]}}$ + NCNN & 91.95  & 93.35 & 94.30 & 95.00\\
    \hline
    \end{tabular}
\end{table}

\subsection{Experiment 4}
Generally, the performances of the speaker recognition system can be effected by the various kernel sizes of the neural network.
One hundred speakers are selected from the FSCMC,
for each speaker, 100 sentences are used as the training data, and another 20 sentences are used as the testing data.
The texts of the speech signals are different between the training data and the testing data.
Four speech features, i.e., ${\pmb{\cal{Y}}}$, ${\pmb{\cal{X}}}$, ${\bf{X}}^{[1]}$, ${\bf{X}}^{[2]}$, are extracted from the sentence of the FSCMC.
If ${{\bf{X}}^{[1]}}$ or ${{\bf{X}}^{[2]}}$ is employed as the input of the NCNN,
the kernel sizes of the NCNN are set to $1 \times 1$, $3 \times 1$, $5 \times 1$ and $7 \times 1$, respectively.
If ${\pmb{\cal{X}}}$ or ${\pmb{\cal{Y}}}$ is used as the input of the NCNN,
the kernel sizes of the NCNN are set to $1 \times 2$, $3 \times 2$, $5 \times 2$ and $7 \times 2$, respectively.
If ${{\bf{Y}}^{[1]}}$ and ${{\bf{Y}}^{[2]}}$ are used as the input of the PCNN-C or PCNN-I,
the kernel sizes of the first convolutional layer of each branch of the PCNN-C or PCNN-I are set to $1 \times 1$, $3 \times 1$, $5 \times 1$ and $7 \times 1$, respectively.
From Table \ref{table pcnn},
for ${{\bf{Y}}^{[1]}}$ + ${{\bf{Y}}^{[2]}}$ + PCNN-I,
the best performances are achieved when the kernel size of the PCNN-I is set to $3 \times 1$.
For ${{\bf{Y}}^{[1]}}$ + ${{\bf{Y}}^{[2]}}$ + PCNN-C,
the best performances are achieved when the kernel size of the PCNN-C is set to $5 \times 1$.
From Table \ref{table multi feature},
the best performances of ${\pmb{\cal{Y}}}$ + NCNN are achieved when the kernel size of the NCNN is set to $3 \times 2$.
The best performances of ${\pmb{\cal{X}}}$ + NCNN are achieved when the kernel size of the NCNN is set to $7 \times 2$.
From Table \ref{table single feature},
for ${{\bf{X}}^{[1]}}$ + NCNN and ${{\bf{X}}^{[2]}}$ + NCNN,
the best performances of the speaker recognition system can be achieved when the kernel size of the NCNN is set to $7 \times 1$.

\section{Conclusion} \label{sec:conclusion}
In this paper, a novel speech feature fusion algorithm is proposed for text-independent speaker recognition.
The TD and the FD features can be extracted from the speaker's speech to build the feature tensor.
Then, the IVA can be utilized to obtain the speaker model, i.e., demixing tensor,
and the IFC matrices of the TD and FD features to remove the redundancies of the same feature type
and enhance the dependence of the different feature types.
The IFC matrices are used as the input of the PCNN to obtain the deep features of the TD and FD features, respectively.
These deep features can be integrated to obtain the fusion feature of the speaker's speech.
The experimental results indicate that the proposed speaker recognition system achieves comparable performances compared with the baseline speaker recognition systems.

\ifCLASSOPTIONcaptionsoff
  \newpage
\fi

\bibliographystyle{IEEEtran}
\bibliography{IEEEabrv,IEEEbib}

%
%

\vfill

\end{document}